\documentstyle[11pt]{article}
\makeatletter
\let\chapter\hid@chapter
\makeatother
\def\lsim{\lower.5ex\hbox{$\; \buildrel < \over \sim \;$}}
\def\gsim{\lower.5ex\hbox{$\; \buildrel > \over \sim \;$}}
\begin{document}
\pagenumbering{arabic}
\title {Monitoring of Sudden Ionospheric Disturbances (SID) from Kolkata (INDIA)}

\author{Sandip K. Chakrabarti$^{1,2}$, K. Acharya$^{2}$, B. Bose$^{2}$, S. Mandal$^2$\\
A. Chatterjee$^{3}$, N.M. Nandi$^3$, S. Pal$^{2}$, R. Khan$^{2,4}$}

\maketitle

\noindent 1. S.N. Bose National Centre for Basic Sciences, JD-Block, Salt Lake, Kolkata 700098\\
\noindent 2. Centre for Space Physics, P-61 Southend Gardens, Kolkata, 700084\\ 
\noindent 3. Centre for Space Physics (Malda Branch), Raja Sarat Bose Rd. 732101\\
\noindent 4. Bidhan Nagar High School, BD block, Salt Lake, Kolkata, 700064\\

\begin{abstract}
We report our first results of monitoring sudden ionospheric disturbances
(SID). We present data taken continuously for two weeks during 20th Sept. 2002
and 4th Oct. 2002. We compare the effects of solar flares of the VLF signal with those  
obtained by GOES Satellite of NASA monitoring in X-rays and found excellent agreement.
\end{abstract}



\noindent Published in Indian Journal of Physics, 2003, 77B, 173

\section{Introduction}

Sudden Ionospheric Disturbances (SID) are caused by ionizing 
sources external to the earth, such as solar and stellar flares,
gamma-ray-bursts, strong X-ray flares near compact objects etc.
By monitoring them, it possible to locate these sources 
and a great deal of knowledge could be obtained about the 
rate at which these types of events occur in nature.

Very Low Frequency (VLF) project of the Centre for Space Physics (CSP)
has been monitoring VLF activities for quite some time. 
During recent Leonid-2002 shower events 
it has detected the peak very distinctly and received 
signals generated by the meteors at 19KHz [1]. In this Rapid Communication,
we present the result of monitoring an Indian Navy Traffic station
at Vijayananarayanam which transmits the signal at 18.2 KHz.
During day time the D-layer of the ionosphere comes down and at night
it goes up (making it ideal for receiving clear signals  at nights) [2].
We not only receive these two effects (sun-set and sun-rise) upon the
signal, we also observe enhanced ionization due to solar flares.
We compare some of these flares with the monitoring of solar X-ray flare
by NASA/GOES satellite which is operative at $0.5$ Ang. to $8.0$ Ang.
The relation among the sun-rise and sun-set effects and airglow (especially
Oxygen Red line at $6300$ Ang.) will be presented in a separate  paper. 
Here, we discuss only the overall behaviour and compare recent solar flare results.

\section{Experimental Setup}

The loop Antenna is  a Gyrator-II type and is 
made up of a square frame of one meter on each side and several turns of 
shielded single core wire is used to receive the signal. The signal is then
amplified and is fed into the audio card of a Pentium-IV computer located inside the
laboratory. The audio signal is 
sampled at $3.2$ times per second. The antenna is tuned at $18.2$KHz, 
the frequency at which VTX3, Indian Navy traffic station at Vijayananarayanam
transmits its signal. It is aligned along the South, South-West direction. 
The magnetic field of the VLF signal induces a current in the antenna
which is then collected as an audio signal.

\section{Results}

Fig. 1 shows  the results of continuous monitoring (barring maintenance
and power failure) the output from September 20th, 2002 to October 4th, 2002.
The days are given in Julian Day and times are in Universal Time (UT).
At the sun rise, the signal drops (around 0h10m each day) 
in a matter of about 15 minutes as the ionization layer 
drops to about 40km about the earth. The sun-rise effect itself has shifted 
by a tens of minutes during our observation since the true sun-rise above the horizon
is shifted. The sun-set (around 37h each day) effect is not very prominent 
and it takes more than one hour and thirty minutes for the D-region to disappear.
Some of the VLF observations from Kolkata were reported earlier at other frequencies [3-4].

In between the sun-rise and the sun-set, several events have been recorded which we
identify as due to the solar flare. For illustration, we consider the 
pair of active flares  occurring on the 29th of September, 2002 (JD 2452545.5+)
in Fig. 2a, one flare occurring on 30th of September, 2002 (JD 2452546.5+)
in Fig. 2c, and one flare occurring on 3rd of October, 2002 (JD 2452549.5+)
in Fig. 2e.  For comparison we  also present GOES satellite data which are
integrated over five minutes obtained from NASA archive [5] (in Figs. 2b, 2d and 2f).
Not only the locations of the peaks match, the profiles also match within 
the error-bars inherent with the audio-card. The duration of the SID is about
one hour or more, consistent with published reports [6] for the same class of flare. 
Very weak signals are masked by our receiving process and cannot be detected. 

\section{Concluding Remarks}

There has been very high degree of solar activity in recent months
as reported by NASA [5]. A large number of sunspots are forming 
and disappearing on the solar disk. These sun-spots produce magnetic flares [2]
which increase the ionization and disturb the constant signal received
by our monitoring station. These Sudden Ionospheric Disturbances (SID)  are being 
monitored at CSP monitoring stations located at Malda and Garia (South Kolkata).
We reported fourteen days of such data obtained at the Kolkata station
which indicated how the ionosphere is disturbed and how long it takes for the
disturbance to subside. We also compared 
the profiles of several `active' flares with the results
obtained by NASA GOES satellite operating to detect X-ray flares on the sun. 
We find excellent agreements in these results, although  the methodologies are
completely different.

\newpage

\centerline{Figure Captions}

\noindent Fig. 1: VLF signals at 18.2KHz is shown for two weeks period stacked
for the sake of comparison. Apart from Sun-rise
and Sun-set effects, the effects of solar flares throughout the day are also observed.
The day numbers (in Julian Day) are written below each curve and the time is in 
Universal Time (UT).

\noindent Fig. 2(a-f):  Comparison of the results obtained by CSP monitoring station
and the results of solar flares obtained  by NASA GOES Satellite operating in X-rays.
In (a), (c) and (e) the CSP data is presented (0.3 sec sampling rate) for flares on 
29th Sept. 2002, 30th Sept. 2002 and 3rd Oct. 2002 while in (b), (d) and (f) the corresponding
NASA observations (five minute average). The times are in Universal Time (UT).

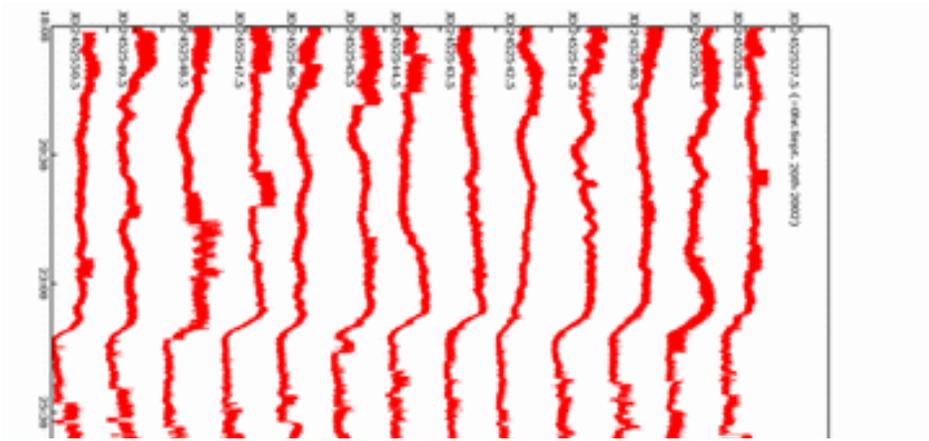
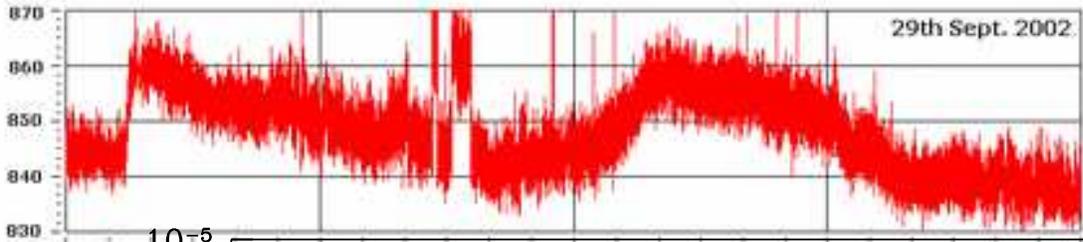
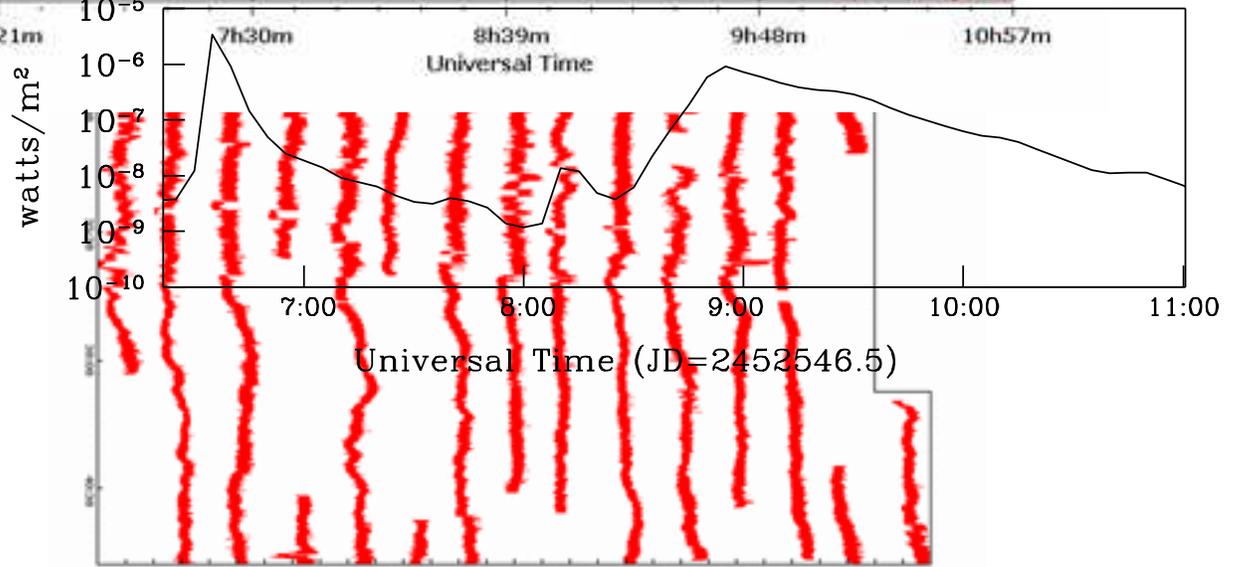
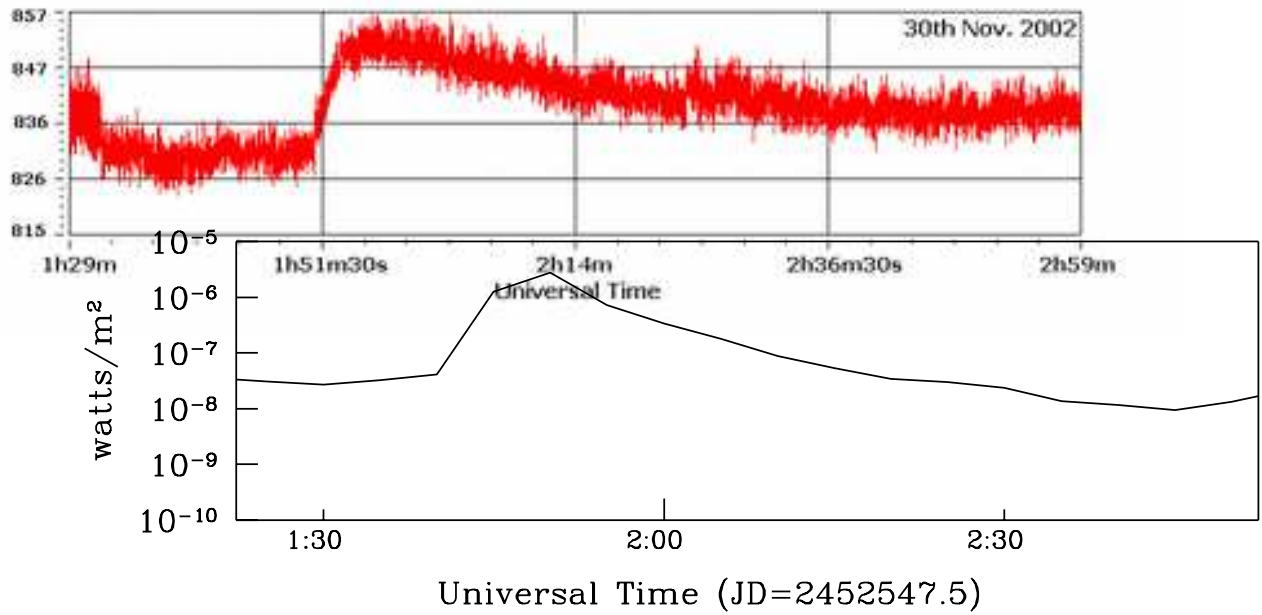
\end{document}